\documentclass[prl,reprint,twocolumn,floatfix,superscriptaddress] {revtex4-1}
\usepackage{amsmath}
\usepackage{graphicx}
\usepackage{lmodern}
\usepackage{amsmath}

\usepackage{color}
\usepackage{amssymb}

\DeclareMathOperator{\sgn}{sgn}

\DeclareMathOperator{\Ree}{Re}

\usepackage{hyperref}

\begin{document}

\title {{Solvable Strong-coupling Quantum Dot Model with a Non-Fermi-liquid Pairing Transition}}
\author{Yuxuan Wang}
\affiliation{Department of Physics, University of Florida, 2001 Museum Rd, Gainesville, FL 32611 USA }
\affiliation{Department of Physics, Stanford University, Stanford, CA 94305, USA}
\begin{abstract}
We show that a random interacting model exhibits solvable non-Fermi liquid behavior and exotic pairing behavior. This model, dubbed as the Yukawa-SYK model, describes the random Yukawa coupling between $M$ quantum dots each hosting $N$ flavors of fermions and $N^2$ bosons that self-tunes to criticality at low energies.
The diagrammatic expansion is controlled by $1/MN$, and the results become exact in a large-$M$, large-$N$ limit. We find that pairing only develops within a region of the $(M,N)$ plane --- even though the  pairing interaction is strongly attractive, the incoherence of the fermions can spoil the forming of Cooper pairs, rendering the system a non-Fermi liquid down to zero temperature. By solving the Eliashberg equation and the renormalization group equation, we  show that  the  transition  into  the pairing phase exhibits Kosterlitz-Thouless quantum-critical behavior.
\end{abstract}
\date{\today}
\maketitle

\noindent\textbf{Introduction.---}
The pairing problem for non-Fermi liquids (nFL) is a fascinating open issue in condensed matter physics~\cite{scal,*scal2,acf,*acs,book1,review3,review4,mack,varma,matsuda,combescot,*Bergmann,*Bergmann2,*ad,*Marsiglio_88,*Marsiglio_91,*Karakozov_91,nick_b,acf,*acs,son,sslee,
subir2,*moon_2,max2,*max_last,raghu_15,mack,scal,*scal2,efetov,steve_sam,Wang2016,tsvelik,vojta,khvesh1,Kotliar2018,*we_last_D,metzner,berg,*berg_2,*berg_3,kotliar,*kotliar2,review3,*tremblay_2,georges,*georges2,*Bergmann,*Bergmann2,*ad,*Marsiglio_88,*Marsiglio_91,*Karakozov_91,acf,*acs,max_last,raghu_15,Wang2016,mandal,first-mats,Kotliar2018,*we_last_D,nonlinear-1,nonlinear-2,phillips-2019}.
In a general context, nFL behavior often occur via electron interactions mediated by gapless bosonic modes~\cite{max,max2,max_last,raghu_15,steve_sam,lawler-barci-fernandez-2006,lawler-fradkin-2007}, rendering the electrons incoherent. Such gapless bosons typically arise in the vicinity of 
a quantum-critical point (QCP) or in gauge theories.
The same interaction is usually also strongly attractive in some pairing-symmetry channel. The fermionic incoherence and the strong attractive interaction compete in determining whether the ground state is superconducting. 
However, the analytical solution of this problem is challenging since there is no natural small parameter in the problem to allow for a controlled calculation for the nFL behavior as well as for the pairing problem. Moreover, the two effects are of comparable strength, lacking a theoretical tuning parameter for the interplay between the nFL and superconductivity. One workaround is to extend the problem to a large-$N$ limit. Within this limit the vertex corrections to the interaction is suppressed by $1/N$, and one can solve for the nFL behavior analytically via self-consistent Schwinger-Dyson equations. Conveniently, the $1/N$ factor also serves as an effective dimensionless coupling constant in the pairing problem. The pairing problem in various large $N$ models for a Fermi surface (FS) coupled to critical bosons have been intensively studied. Interestingly, in a class of these models~\cite{acf, raghu_15, wang-wang-torroba}, as a function of $N$, the system at $T=0$ can either be in a pairing phase, or remain at the normal state, separated by a quantum-critical point. The latter situation is particularly striking --- contrary to BCS theory where even an infinitesimal attractive interaction drives a Fermi liquid to a superconducting state, the incoherence of the nFL state destroys superconductivity, \emph{even if} the attractive pairing interaction is strong. However, certain large-$N$ extensions become uncontrolled in two spatial dimensions~\cite{sung-sik-09,max,max2}, and the quantum critical point for pairing can only be accessed for FS's in fractional spatial dimensions $d=3-\epsilon$ ($0<\epsilon<1$)~\cite{raghu_15,mandal}, the physical meaning and effective realization of which are unclear. {It naturally raises the question whether the pairing QCP may be an artifact of the fractional spatial dimensions.}

Recent years have witnessed a remarkable revival of interest in the Sachdev-Ye-Kitaev (SYK) models~\cite{SY, K, SYK2, SYK3},  due to its property of maximal quantum chaos~\cite{maxchaos} and its connection with quantum black holes. These models describe random four-fermion interactions within a quantum dot of $N$ fermionic particles. Despite being strongly interacting, these models can be solved in the large-$N$ limit and exhibit nFL behavior~\cite{SY, balents-2017, debanjan-prx, xu-2019, franz-2019}. In the pairing problem for the SYK model as well as its lattice variants, the nFL was found to be generally unstable to pairing~\cite{cenke-syk-1,cenke-syk-2,patel-kim-2018, zhao-prx-2019, chowdhury-berg-2019,gnezdilov-2019} in the presence of a small attractive interaction.

In this Letter we study a new solvable random interacting model with more exotic nFL pairing behaviors. This model describes $M$ quantum dots each with $N$ flavors of fermions. The fermions are coupled by a random Yukawa term to a matrix boson with a generic bare mass, and we thus refer to this model as the Yukawa-SYK model. We show that the coupling with the fermions makes the boson critical, \emph{independent} of its bare mass.
The critical behavior is similar to that recently obtained for a model proposed in Ref.~\onlinecite{patel2018}, where instead of Yukawa coupling a minimal coupling to a compact dynamical $U(1)$ gauge boson was introduced. However, we show that there the compactness of the gauge field actually confines the fermions and spoils the nFL behavior in that model.~\cite{suppl, patel-private}
Like the SYK model, we show that this model is solvable in the large-$N,M$ limit and exhibit nFL behavior. As one varies the ratio $N/M$, the exponent of the conformal fermionic self-energy interploates between 0 (same as in a noninteracting disordered electron system) and 1/2 (same as in the SYK model). {We also determine the nonuniversal overall scale of the nFL self-energy, by matching the UV and IR properties.}
Remarkably, the large-$N,M$ limit also allows for an analytical solution of the pairing gap equation in the nFL regime. By solving the gap equation as an integral equation we show that the pairing phase only develops 
for 
\begin{equation}
M\leq \sqrt{2N}, M,N\to \infty.
\end{equation}
Outside this range,  the system remains a nFL down to zero temperature, \emph{even if} the attractive interaction is singularly strong. Moreover, the pairing gap near the pairing QCP $M_{cr}=\sqrt{2N_{cr}}$ follows a Kosterlitz-Thouless (KT) scaling form and represents an \emph{infinite-order} phase transition [see Eq.~\eqref{KT}], similar to that found in Ref.~\onlinecite{raghu_15} in fractional dimensions. This KT scaling near the pairing QCP can be understood as coming from the annihilation of two renormalization group (RG) fixed points~\cite{kaplan2009, holoQM}, which we show in Ref.~\onlinecite{suppl}.

\noindent\textbf{The Yukawa-SYK model.---}\label{sec:model}We consider a random interacting model {of $M$ quantum dots  each with $N$ flavors of fermions $(c)$ coupled with a critical boson $(\phi)$ through a random Yukawa term.} %gauged  random hopping between electrons within and among quantum dots (clusters), where the gauge field couples to inter-cluster hoppings (see Fig.~\ref{cartoon}).  
The Hamiltonian is given by
\begin{equation}
{H_0}=\frac{i}{(MN)^{1/2}}\sum_{ij,\alpha\beta}^{N,M} t_{\alpha\beta}\phi_{ij}c^\dagger_{i\alpha}c_{j\beta} + \frac{1}{2}\sum_{ij}^{N}\left (\pi_{ij}^2 +{m_0^2}\phi_{ij}^2\right ).
\label{ham}
\end{equation}
{where $\alpha,\beta \in (1,M)$ are indices for the quantum dots and $i,j\in (1,N)$ are $SO(N)$ flavor indices within a cluster.} When taking the large $N,M$ limit one {can vary} the ratio $M/N$. %For the purpose of singlet pairing, the fermions have spin-$1/2$, and the summation over spinor indices is implicit.
%We also include a bare-mass counterterm $\delta H=\Lambda\sum_{ij}\phi_{ij}^2/2$, and we discuss the value of $\Lambda$ later.
 Here $\pi_{ij}= \dot{\phi}_{ij}/g$ is the canonical momentum of the boson field $\phi_{ij}$. As we will see, the infrared (IR) dynamics of the boson field completely comes from its coupling with the fermions. However its bare dynamics is important for fixing the energy scale of low-energy dynamics. {Note that the $i$ factor in the first term and Hermiticity of the Hamiltonian indicates that {$\phi_{ij}\equiv -\phi_{ji}$ (assuming $t_{\alpha\beta}=t_{\beta\alpha}$).} This will be important for the pairing problem.}  The random coupling amplitudes satisfy a Gaussian distribution
where 
\begin{equation}
\langle t_{\alpha\beta} t_{\alpha'\beta'}\rangle = \omega_0^3{(\delta_{\alpha\alpha'}\delta_{\beta\beta'}+\delta_{\alpha\beta'}\delta_{\beta\alpha'}).}
\end{equation}
{This model in Eq.~\eqref{ham} is similar to that recently studied in Ref.~\onlinecite{patel2018}, which has a similar $N,M$ assignment. There the fermions are randomly coupled to a compact gauge field and the authors obtained a nFL behavior at low-energies, much similar to our results below. However, we show in Ref.~\onlinecite{suppl} that in the compact gauge theory fluctuations of the gauge field actually drives the system into a  confined phase~\cite{patel-private}}. 

{Despite being a toy model far from realistically describing known condensed-matter systems,  we shall see  it exhibits interesting low-energy behaviors that are universal and independent of details, just like the SYK models~\cite{SY, K, SYK2, SYK3}. (For example, one can construct an alternative model free of randomness that behaves essentially the same~\cite{suppl}). Further, we show below the normal state and the pairing state [Eqs.~(\ref{eq2}, \ref{BS})] are described by the Eliashberg equations widely adopted to describe electron-phonon superconductors and quantum-critical superconductors.~\cite{Wang2016} From this perspective the intricate structure of the Hamiltonian is merely a tool to make the various usual theoretical simplifications quantitatively controlled, much the same way as the usual large-$N$ approximation.}

\begin{figure}[t]
\includegraphics[width=0.7\columnwidth]{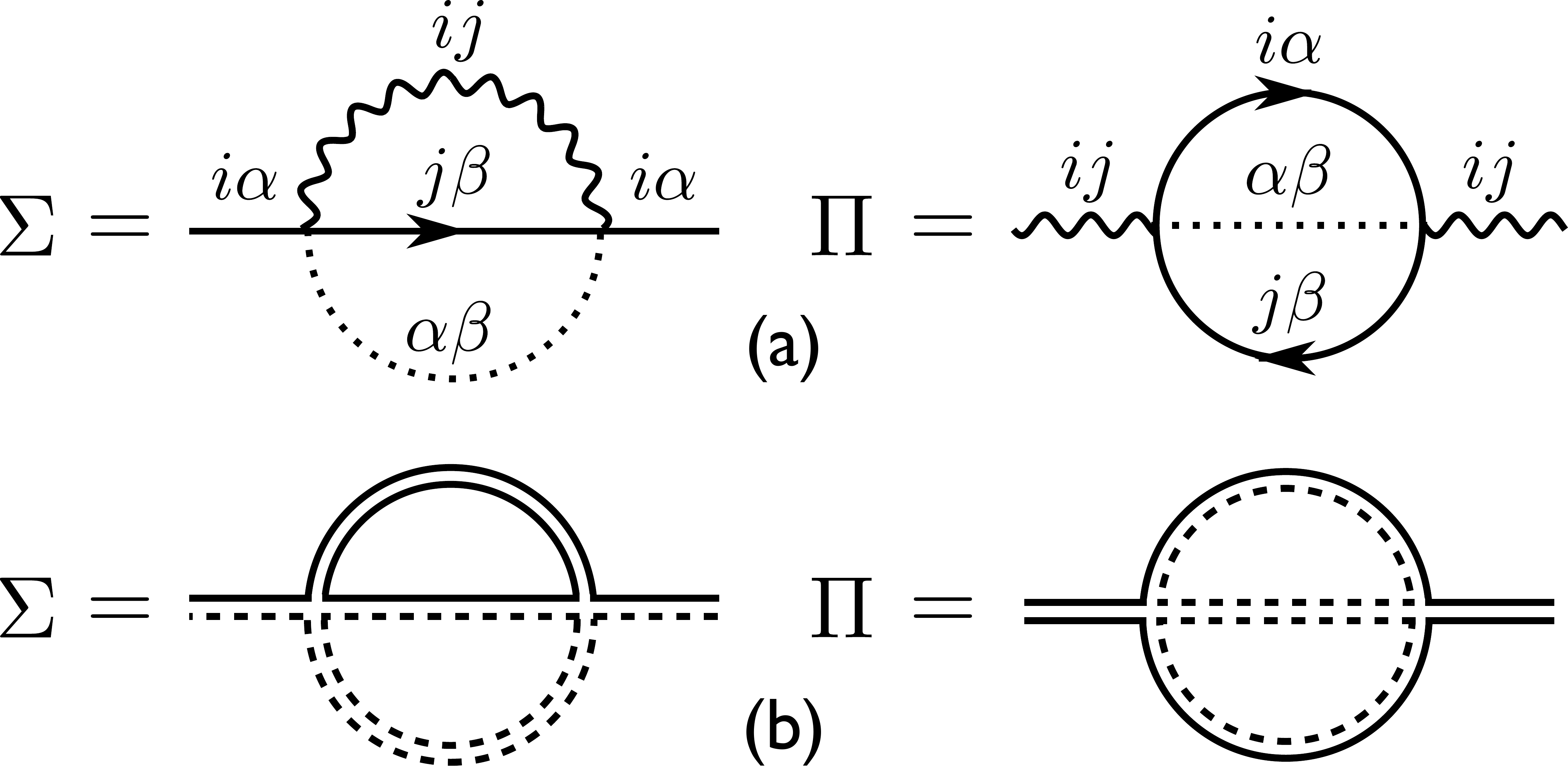}
\caption{{(a) The Feynman diagrams corresponding to the self-energies of the bosons and fermions. (b) The large-$N,M$ behavior of the diagrams can be tracked using a  double-line formalism. The solid line represents indices $i,j$ and the dashed lines $\alpha,\beta$.}}
\label{SE}
\end{figure}

\noindent\textbf{Normal state analysis.---}\label{sec:norm}To leading order in $1/(NM)$, the self-consistent Schwinger-Dyson equations for the bosonic and fermionic self-energies {(in Euclidean time at $T=0$)} are
\begin{align}
\label{eq2}
\Sigma(\omega) =& \omega_0^3 \int \frac{d\Omega}{2\pi} D(\Omega)G(\omega+\Omega),\\
\Pi(\Omega) =& 2\omega_0^3\frac{M}{N}\int\frac{d\omega}{2\pi}G(\omega-\Omega/2)G(\omega+\Omega/2)\nonumber
\end{align}
where $D(\Omega)= 1/\left [\Omega^2 +m_0^2+\Pi(\Omega)\right ]$, and $G(\omega)=-1/\left [i\omega+\Sigma(\omega)\right ]$.   We show the corresponding diagrams in Fig.~\ref{SE}. {The fact that these diagrams are to leading order in $1/(NM)$ can be seen using a double-line formalism, keeping track of both $i,j$ (solid line in Fig.~\ref{SE} (b)) and $\alpha,\beta$ (dashed line) indices; all other diagrams are suppressed by $1/(NM)$.} 

{As a first attempt, we evaluate the diagrams with bare propagators, and perturbation theory immediately fails: the integral}
\begin{align}
\Pi(0)= 2\omega_0^3\frac{M}{N}\int\frac{d\Omega}{2\pi} \frac{1}{(i\Omega)^2}
\label{eq:bareb}
\end{align}
{is strongly  infrared divergent, which would drive the bosonic mass $m^2=m_0^2+\Pi(0)$ towards negative. In this situation, the boson usually condenses described by a mean-field theory. Nevertheless, we are in 0d and fluctuation effects typically destroy any order. Large-$M$ limits, $M$ being the number of fermions, have often been invoked to suppress flucutation by $1/M$ and can stabilize order even in 0d. However, as for each boson there are $M/N$ fermions, one can check that in our case fluctuation effects of the boson are $O(N/M)$, in general not small. The fact that the bosons and the fermions have comparable amount of fluctuations spoils a simple mean-field solution. Instead, such a balance of fluctuation effects gives rise to a new, critical solution for the ground state. In this state, the divergence in the bare bubble \eqref{eq:bareb} is cut off at the low energy, the bosonic mass $m^2$ gets renormalized to zero, and makes the fermions a nFL at low-energies, which is in turn consistent with the cutoff in the bare fermion bubble.}

We analyze the solution for which at low energies, $\Pi(0)= m_0^2$ (criticality), $\Omega^2/g\ll \Pi(\Omega)-\Pi(0)$ and $\omega\ll \Sigma(\omega)$.
We can take the conformal nFL ansatz for the self energies {(at $T=0$)}: 
\begin{align}
\Sigma(\omega) =& iA \omega_0^{1-x} |\omega|^x \sgn(\omega),~\nonumber\\
\tilde \Pi(\Omega)\equiv & \Pi(\Omega)-\Pi(0) =B \omega_0^{1+2x} |\Omega|^{1-2x}.
\label{eq:5}
\end{align} 
%%%%%%
and the power $x$ can be determined by using the following integrals
\begin{align}
&\int \frac{d\Omega\sgn(\omega-\Omega)}{|\Omega|^{1-2x}\,|\omega-\Omega|^x} = -\frac{\Gamma^2(-x)}{2\Gamma(-2x)}\,|\omega|^x\sgn(\omega)\nonumber\\
%&\equiv 2\pi\alpha|\omega|^x\sgn(\omega),\nonumber\\
&\int \frac{d\omega\, \sgn(\omega+\Omega/2)\,\sgn(\omega-\Omega/2)}{|\omega+\Omega/2|^x\,|\omega-\Omega/2|^x}
\nonumber\\
& = |\Omega|^{1-2x}\,\frac{\Gamma^2(-x)}{2\Gamma(-2x)}\, \frac{1+\sec{\pi x}}{1/x-2} +{\rm div} %&\equiv -2\pi \tilde \alpha|\Omega|^{1-2x} + {\rm div},
\label{int}
\end{align}
where ``div" stands for a divergent constant to be regularized by high-energy physics. Defining
$\alpha(x) = -{\Gamma^2(-x)}/[{4\pi\Gamma(-2x)}]
$,
self-consistency of Eq.~\eqref{eq2} requires~\cite{patel2018} %{(put back the details, but be extra careful about the signs!)}
\begin{align}
A^2B=\alpha(x),&~\frac{2M}{N} = \frac{1/x-2}{1+\sec(\pi x)}.
\label{MN}
\end{align}
We take $0<x<1/2$ to guarantee $B>0$, a requirement of causality~\footnote{Note that the equation for $x$ is similar to that of the supersymmetric SYK model at $\mathcal{N}=1$~\cite{susy-syk}. However there the restriction $B>0$ leads to $1/2<x<1$, and the solution for $x$ yields $x=2/3$, consistent with supersymmetry. There the $0<x<1/2$ solution  breaks the supersymmetry and is unstable, opposite to the present model.}.
For $N\ll M$, $x\to0$ and for $N\gg M$, $x\to 1/2$. 

{Low-energy physics alone does not determine  the prefactors $A$ and $B$, which is quite different from finite-dimensional models where the feedback effect from the nFL to the critical boson is negligible.~\cite{acs}  To fix them, one needs to examine the system behavior at high energies. For Eq.~\eqref{eq:5} to hold, self-energies should be dominant over the bare terms in the propagators. This requires $\omega,\Omega \ll \bar\omega \equiv \min (\omega_F,\omega_B)$, where from Eq.~\eqref{eq:5}  $\omega_F \equiv \omega_0 A^{1/(1-x)}$ and $\omega_B \equiv \omega_0 B^{1/(1-2x)}$. 

 If $\omega_F\ll \omega_B$, $\omega_F$ serves as an effective infrared cutoff for the otherwise-divergent integral in Eq.~\eqref{eq:bareb}. From
$m_0^2-\int_{\omega_F}{\omega_0^3d\omega}/{\omega^2} = 0$ we get 
$\omega_F \sim {\omega_0^3}/{m_0^2}.$
The requirement $A^2B = \alpha(x)$ yields
\begin{align}
\Sigma(\omega) =& i c\ \omega_{F}^{1-x}|\omega|^x \sgn(\omega), \nonumber\\
\tilde\Pi(\Omega) =&c^{{-2}}\alpha(x) m_0^2 |\Omega/\omega_{F}|^{1-2x}~~(\omega_F= {\omega_0^3}/{m_0^2}.)
\label{ansatz3}
\end{align}
where $c$ is a nonuniversal $O(1)$ constant. From this $\omega_B^{1+2x} = m_0^3/\omega^{1-2x}$, and the $\omega_B\gg \omega_F$ translates to $\omega_0\ll m_0$, which we  denote as the weak-coupling region. 

If $\omega_B\ll \omega_F$, $\Sigma(\omega)$ takes a different form for $\omega\gg \omega_B$. In this regime, the fermions have  much higher energies than the typical boson energy $\omega_B$, and the latter behaves like disorder:
\begin{align}
\Sigma(\omega)&=-\frac{\omega_0^3}{\Sigma(\omega)}\int\frac{d\Omega}{\Omega^2 + \tilde \Pi(\Omega)} \sim\sqrt{\frac{\omega_0^3}{\omega_B}}\sgn(\omega),
\label{eq:90}
\end{align}
The new scale $\sqrt{\omega_0^3/\omega_B}$ serves to cut off the divergence in Eq.~\eqref{eq:bareb}, and thus $\omega_B\sim {m_0^4}/{\omega_0^3}$. Below the $\omega_B$ scale, $\Sigma,\Pi$  restore the forms in Eq.~\eqref{eq:5}. Using self-consistency and $\omega_B = \omega_0B^{1/(1-2x)}$
\begin{align}
\Sigma(\omega) = & id {\omega_{0}^{3/2}}{\omega_B^{-1/2-x}} |\omega|^x \sgn(\omega), \nonumber\\
\tilde\Pi(\Omega) =&d^{-2} \alpha(x){\omega_B^{1+2x}|\Omega|^{1-2x}}~~(\omega_B={m_0^4}/{\omega_0^3})
\label{eq:100}
\end{align}
where $d=O(1)$, and we see that at $\omega\sim\omega_B$, $\tilde\Pi(\Omega) = \Omega^2$ by definition and that  $\Sigma(\omega)$ smoothly crosses over between Eqs.~(\ref{eq:90}, \ref{eq:100}). The condition $\omega_B\ll \omega_F$ implies $\omega_0\gg m_0$, which we denote as the strong-coupling regime.

Remarkably, we have shown that for arbitrary values of $m_0$, the system behavior  \emph{self-tunes} to {criticality} (see also Ref.~\onlinecite{esterlis}), in sharp contrast of how similar systems behave at finite dimensions. We discuss the self-energies at higher frequencies in both regimes in Ref.~\onlinecite{suppl}.

\begin{figure}[t]
\includegraphics[width=0.6\columnwidth]{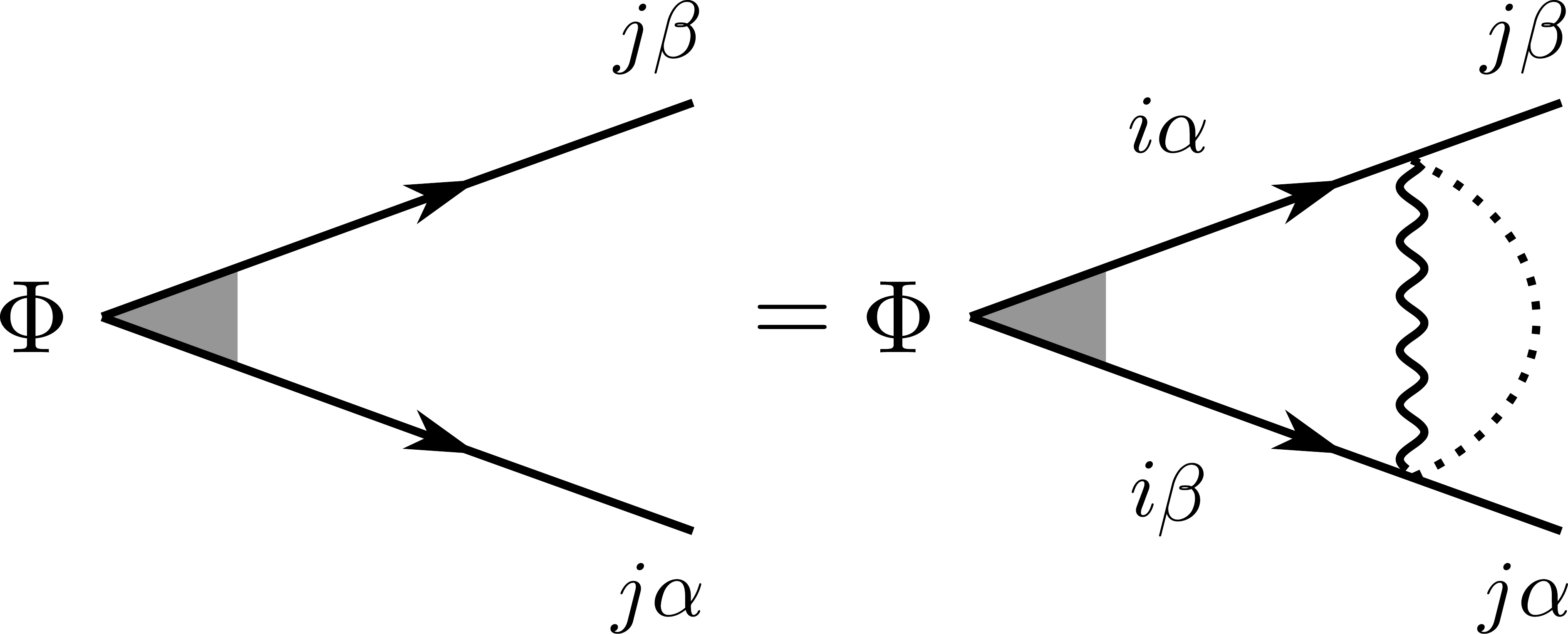}
\caption{The gap equation for the inter-dot pairing. One can show that the interaction for all other channels is repulsive.}
\label{diag2}
\end{figure}

\noindent\textbf{The Pairing problem.---}
\label{sec:pair}
The diagram for the gap equation of  the pairing problem are shown in Fig.~\ref{diag2}. %The gap equation is formulated self-consistently --- the shaded triangle vertex on the right hand side is the pairing order parameter $\Phi$ itself.
%Iterating the expression for $\Phi$ leads to the familiar summation over ladder diagrams.
 We consider {inter-dot} pairing with  $\alpha\neq \beta$. Defining the pairing vertex {$\sim \Phi_{\alpha\beta}c^\dagger_{i\alpha}c^\dagger_{i\beta}$, fermion statistics requires $\Phi_{\alpha\beta}=-\Phi_{\beta\alpha}\equiv \Phi$.~\footnote{{We have assumed  phase coherence between Cooper pairs with different $\{\alpha,\beta\}$. This can be ensured by turning on an arbitrarily small attractive pair-hopping interaction.}} Indeed the interaction mediated by exchanging $\phi_{ij}$ is attractive in this channel [note the $i$ factor in Eq.~\eqref{ham}].} The Eliashberg equation for pairing is given by
{\begin{align}
\label{BS}
\!\!\Phi(\omega) =& \frac{t^2T}{M}\sum_{\Omega}D(\Omega)|G(\omega+\Omega)|^2\Phi(\omega+\Omega).
\end{align}}Compared with the normal state analysis, the right hand side of \eqref{BS} is suppressed by $1/M${, as only the internal flavor index $i$ is summed over.} {Restricting to even-frequency pairing, one can verify that in our model the interactions for \emph{all} other pairing channels are repulsive.} 

In this work we focus on the pairing problem at $T=0$. To find the pairing gap up to an $O(1)$ coefficient, it suffices to solve the linearized gap equation in the presence of an infrared cutoff $\Delta$, roughly the magnitude of the pairing gap $\sim\omega\Phi(\omega)/\left [\omega + \Sigma(\omega)\right ]$. We place an effective UV cutoff at the nFL energy scale $\bar\omega=\min (\omega_F, \omega_B)$, and pairing comes from physics below this scale. The gap equation becomes
\begin{equation}
{\Phi(\omega) = \frac{2}{\alpha(x) M}\int^{\bar\omega}_{\Delta} \frac{d\omega'}{2\pi} \frac{ \Phi(\omega')}{|\omega-\omega'|^{1-2x}|\omega'|^{2x}}.}
\label{eq21}
\end{equation}
Note that the nonuniversal $c,d$ dependence in Eqs.~(\ref{ansatz3}, \ref{eq:100}) has disappeared.

In sharp contrast with the BCS pairing, the integral in \eqref{eq21} does not explicitly contain a logarithmic IR divergence~\cite{acf}, at least with a constant $\Phi$. In fact Eq.~\eqref{eq21} is very similar to the gap equation in  quantum-critical pairing problems~\cite{acf, first-mats, wang-wang-torroba}, and to see the pairing instability we  need to solve the  full integral equation. We extend the domain of $\omega'$ to $(0,\infty)$ and split the integral on the right hand side:
\begin{align}
\Phi(\omega) =& \frac{2}{\alpha M}\left [\int_{0}^{\infty}-\int_{\omega_0}^{\infty} -\int_{0}^\Delta\right ]\frac{d\omega'}{2\pi}\frac{\Phi(\omega')}{|\omega-\omega'|^{1-2x}|\omega'|^{2x}}.
\label{eq260}
\end{align}
Note that the first integral alone can be matched with the left hand side by using a power law ansatz 
\begin{align}
\Phi(\omega)=& \omega^{-y},~~{\pi\alpha(x)M}=\int_{0}^\infty\frac{du}{|1-u|^{1-2x}|u|^{2x+y}}.
\label{eq26}
\end{align}
 where $0<\Ree y<1-2x$. %, and $y$ satisfies
%{ \begin{equation}
%$\frac{\alpha(x)M}{2}=\int_{0}^\infty \frac{du}{2\pi}\frac{1}{|1-u|^{1-2x}|u|^{2x+y}}.$
%\label{eq26}
%\end{equation}}
We then need to make sure the other two integrals in \eqref{eq260} vanish for \emph{external} frequencies $\Delta\ll \omega \ll \omega_0$.
However, they could only vanish if $\Phi(\omega')$ is oscillatory in those intervals. This requires $y$  to be complex.~\cite{acf,first-mats} 

One can show from Eq.~\eqref{eq26} that for large $\alpha(x)M$, $y$ takes real values, and a complex $y$ is only possible for small enough $\alpha(x)M$. The critical value for $\alpha(x) M$ is given by the minimum value of the right hand side of \eqref{eq26} for a real $y$, which is reached at $y=(1-2x)/2$. We then find the critical value to be
{\begin{equation}
{\alpha(x_{cr})M_{cr}}= \frac{2^{2x_{cr}}\Gamma(\frac{1}{2}-x_{cr})\Gamma(x_{cr})}{{\pi}^{3/2}}. 
\label{eq29}
\end{equation}}Together with Eq.~\eqref{MN}, this gives a line of critical values  $(M_{cr}, N_{cr})$. We will see that this region denotes the pairing phase at $T=0$. 
In the limit $N, M \to\infty$ where our results become exact, one can verify that the critical values satisfy 
$
M_{cr}=\sqrt{2N_{cr}}\ll N_{cr},
$
and for $M>\sqrt{2N}$ the system remains a nFL down to zero temperature.}
 
For $M\leq\sqrt{2N}$, the solution for $y$ is complex,  $y = (1-2x)/2 \pm i\beta$. Near the critical pairs $(N,M)_{cr}$, $\beta$ scales as
%\begin{equation}
$\beta\propto  \sqrt{ \lambda(N,M)-\lambda_{cr}(N,M) },
$ %\end{equation}
where $\lambda(N,M)\equiv 1/(\alpha M)$. %(We remind that $\alpha(x)$ is ultimately a function of $M,N$.) 
The power-law ansatz can be rewritten as 
\begin{align}
\Phi(\omega)=\omega^{-(1-2x)/2} \cos[\beta \log \omega + \phi],
\label{sing}
\end{align}
where $\phi$ is a free parameter.
With this form for $\Phi(\omega')$, requiring the second and third integrals in \eqref{eq260} to vanish determines the value of $\phi$ and $\Delta$. % the SC gap $\Delta$:
%\begin{align}
%&\int_{0}^{\Delta} \frac{d\omega' \cos[\beta \log \omega'+\phi]}{|\omega-\omega'|^{1-2x}|\omega'|^{x+1/2}} \nonumber\\&\approx \frac{1}{|\omega|^{1-2x}}\int_{0}^{\Delta} \frac{d\omega' \cos[\beta \log \omega'+\phi]}{|\omega'|^{x+1/2}} = 0, \nonumber\\
%&\int_{g}^{\infty} \frac{d\omega' \cos[\beta \log \omega'+\phi]}{|\omega-\omega'|^{1-2x}|\omega'|^{x+1/2}} \nonumber\\&\approx \int_{g}^{\infty} \frac{d\omega' \cos[\beta \log \omega'+\phi]}{|\omega'|^{3/2-x}} = 0,
%\nonumber\end{align}
%%where in the second step of each line we have used the fact that the external frequency $\Delta\ll\omega \ll g$.  
Using the fact that the external frequency $\Delta\ll\omega\ll \omega_0$, 
we obtain
\begin{align*}
\tan(\beta \log \Delta +\phi) = \frac{2\beta}{1-2x}, \tan(\beta \log \bar\omega +\phi) = -\frac{2\beta}{1-2x}.
\end{align*}
The solution of $\Phi(\omega)$ that does not change sign within $(\Delta,\omega_0)$ maximizes the condensation energy. Requiring this we get at small $\beta$,
\begin{align}
\!\!\!\Delta \sim \bar\omega\exp{\left (-\frac{\pi}{\beta}\right )} ={ \bar\omega\exp\left(-\frac{\gamma M^{3/4}N^{-1/4}}{\sqrt{2-M^2/N}}\right)},
\label{KT}
\end{align}
where $\gamma$ is an $O(1)$ number. {Indeed, this region with $M\leq\sqrt{2N}$ corresponds to a pairing phase.}
Furthermore, we see that near the pairing QCP $\Delta$ onsets via an \emph{infinite-order} phase transition similar to a KT transition.~\cite{altland}  %, i.e.,
%\begin{equation}.
%\Delta \sim g\exp\left (-\frac{\#}{\sqrt{f(M,N)_{\rm cr}-f(M,N)}}\right ).
%\end{equation}
{The connection to the KT transition can be made clear in an RG framework:} this exotic KT scaling of the pairing QCP  comes from the merger of two fixed points~\cite{raghu_15}, which we explain in Ref.~\onlinecite{suppl}. This scaling
was also found for quantum-critical pairing models~\cite{acf,raghu_15}, as well as in some holographic models~\cite{kaplan2009, holoQM}.  %However such a behavior the large-$N$ quantum-critical pairing models requires a fractional spatial dimension.
% It would certainly be interesting to see if the present model has a holographic dual.
%In summary, we have found a scenario where even though the pairing interaction is not only attractive but also very strong, the fermionic incoherence in a nFL makes the electrons unable to pair down to $T=0$. There exist a line of quantum critical points for pairing with KT scaling behavior.
%The pairing behavior at finite temperatures is qualitatively similar with some subtleties, as we discuss in \cite{sm}. 
This is the main result of this work.  %Due to the gapless gauge interaction, the pairing problem for this nFL model is fundamentally different from that in the BCS theory. Both the pairing vertex and the interaction have nontrivial frequency dependence, and one has to solve the full integral equation. As a physical consequence, the nFL behavior can be stabilized against pairing down to zero temperature.

{It is again important to address whether the above mean-field result for pairing is destroyed by fluctuations of the order parameter $\Phi_{\alpha\beta}$. In the present case there are $M^2$ boson species coupled to $NM$ fermion species. Therefore the flucutation effects of $\Phi_{\alpha\beta}$ beyond its mean-field theory is $O(M/N)$ (compare with the $O(N/M)$ obtained previously for the fluctuation effects of $\phi_{ij}$). In the region $M\leq \sqrt{2N}$, such effects are suppressed.  }

\noindent\textbf{Conclusion.---} \label{sec:conc}
The interplay between nFL and pairing has been a long standing open issue due to the lack of a natural control parameter. We have shown in an exactly solvable large-$N$ random interacting model that the opposite tendencies of fermionic incoherence and strong attraction from the \emph{same} interaction lead to remarkable consequences --- for a large range of $(N,M)$, the nFL behavior completely spoils the Cooper pairing, despite the pairing interaction mediated by critical bosons is singularly strong. Only for some values that asymptote to $M_{cr}\leq\sqrt{2N_{cr}}$, the system enters a pairing phase. 
By solving the Eliashberg equation, we have shown that the $T=0$ critical point between the pairing phase and the nFL phase exhibit a KT scaling behavior. Unlike previous models exhibiting this behavior that requires a fractional spatial dimension, the present model has a well-defined base manifold. It will be interesting to explore its experimental and numerical realizations.

An interesting question is the quantum chaotic behavior across the pairing QCP. The low-energy conformal invariance indicates that the nFL state should saturate the upper bound of Lyapunov exponent $\lambda_L$ and is dual to a quantum black hole~\cite{maxchaos,kim-cao-altman-2019}, just like the SYK model. If so, it will be interesting to see how the coefficient of $\lambda_L$ behave across the pairing QCP.  Qualitatively, we expect $\lambda_L$ to drop due to the formation of the condensate. We postpone a full analysis of $\lambda_L$ to future studies. {Another interesting question for future work is the interpretation of the KT scaling of the superconducting transition from the perspective of the 2d classical bulk theory.}

\begin{acknowledgments}

I thank I.~Esterlis, S.~Kivelson, A.~Patel, S.~Raghu, Y.~Schattner, J.~Schmalian, G.~Torroba, H.~Wang, and especially A.~V.~Chubukov for stimulating discussions. This research was initiated at the Aspen Center for Physics, supported by NSF PHY-1066293. This work was funded by the Gordon and Betty Moore Foundation’s EPiQS Initiative through Grant GBMF4302 and GBMF8686.

\emph{Note added:} After the completion of this work, I learned about an independent study of random \emph{real} coupling between the fermions and phonons by Ilya Esterlis and J\"org Schmalian~\cite{esterlis}.  
Without the $i$ factor in the Yukawa coupling, their interaction is attractive in the intra-flavor pairing channel and not suppressed by $1/N$. Hence pairing already develops at $N=\infty$. Our normal state results agree. I am grateful to them for sharing  their unpublished work with me. After the submission of this work, a followup mini-review of their work has appeared~\cite{esterlis-2}.

\end{acknowledgments}

\bibliography{pairingQCP.bib}

\end{document}